\newif{\ifjournal}
\newif{\ifarticle}
\journalfalse
\articlefalse

\newlength{\fsize}

\ifjournal
  \documentclass{aa}
  \usepackage{graphicx,times}
\else
  \ifarticle
    \documentclass[12pt]{article}
    \usepackage{a4wide,amssymb,graphicx,mathptm}
  \else
    \documentclass{paper}
  \fi
  \newcommand{\la}{\lesssim}
  \newcommand{\ga}{\gtrsim}
\fi

\renewcommand{\d}{\mathrm{d}}

\begin{document}

\ifjournal
  \title{Lensing Sunyaev-Zel'dovich Clusters}
  \author{Matthias Bartelmann}
  \institute{Max-Planck-Institut f\"ur Astrophysik, P.O.~Box 1317,
    D--85741 Garching, Germany}
  \date{\today}

  \authorrunning{M. Bartelmann}
  \titlerunning{Lensing Sunyaev-Zel'dovich Clusters}
  \thesaurus{11.03.1 -- 12.03.1 -- 12.07.1}
  \maketitle
\else
  \title{Lensing Sunyaev-Zel'dovich Clusters}
  \author{Matthias Bartelmann\\
    Max-Planck-Institut f\"ur Astrophysik,
    \ifarticle\\\fi P.O.~Box 1317, D--85741 Garching, Germany}
  \date{\em Astronomy \& Astrophysics, submitted}
  \ifarticle\maketitle\fi
\fi

\begin{abstract}

Full-sky microwave surveys like the upcoming {\em Planck\/} satellite
mission will detect of order $10^4$ galaxy clusters through their
thermal Sunyaev-Zel'dovich effect. I investigate the properties of the
gravitationally lensing subsample of these clusters. The main results
are: (1) The combined sample comprises $\ga70\%$ of the complete
sample. (2) It is confined to redshifts $0.2\pm0.1$, and to masses
$(5\pm3)\times10^{14}\,M_\odot$. (3) Using a particular measure for
the weak lensing effect, viz.~the aperture mass, cluster masses can be
determined with a relative accuracy of $\sim20\%$ if their density
profile is known. Consequently, the mass function of the combined
sample can accurately be measured. (4) For low-density universes, I
predict a sharp peak in the measured (aperture) mass function near
$5\times10^{14}\,M_\odot$ and explain its origin, showing that the
peak will be absent in high-density universes. (5) The location of the
peak and the exponential decrease of the mass function on its
high-mass side will allow the determination of the amplitude of the
dark-matter power spectrum on the cluster scale and the baryon
fraction in clusters, and constrain the thermal history of the
intracluster gas.

\ifjournal
\fi

\end{abstract}

\ifjournal\else\ifarticle\else\maketitle\fi\fi

\section{Introduction\label{sec:1}}

Upcoming full-sky surveys in the microwave regime, like the {\em
Planck\/} mission (Bersanelli et al. 1996), with angular resolutions
of order $(5-10)'$ and sensitivities of micro-Kelvins, will detect
thousands of galaxy clusters through their thermal Sunyaev-Zel'dovich
effect. This effect arises because hot thermal electrons in the
intracluster plasma Compton-upscatter the much colder
microwave-background photons, re-distributing them from the
low-frequency part of the spectrum below 218~GHz to the high-frequency
part (Sunyaev \& Zel'dovich 1972, 1980; Rephaeli 1995; Birkinshaw
1999).

At angular resolutions of a few arc minutes, most clusters will be
barely resolved or unresolved (e.g.~Aghanim et al. 1997; Hobson et
al. 1998). Detailed cluster studies with the CMB survey data alone
will therefore not be possible. In combination with optical, radio, or
X-ray follow-up observations, however, the rich cluster samples
detected in the microwave regime can be used to extract a wealth of
cosmological information.

Promising additional information can be gained exploiting the
gravitational lensing effects of the Sunyaev-Zel'dovich cluster
sample. While strong lensing effects depend on the accurate alignment
of randomly distributed background sources with respect to the cluster
lenses, weak lensing is much less affected by stochastic effects. The
question therefore arises what kind of information can be extracted
from weak-lensing follow-up observations of the cluster sample
produced by CMB missions such as {\em Planck\/}.

In this paper, I discuss in Sect.~\ref{sec:2} the detection of
clusters by {\em Planck\/} and the properties of the resulting cluster
sample. In Sect.~\ref{sec:3}, weak gravitational lensing is introduced
and applied to the cluster sample. The mass distribution of the
combined sample is discussed in Sect.~\ref{sec:4}, and it is shown how
it can be measured. The main results are summarised and discussed in
Sect.~\ref{sec:5}.

\section{Planck's cluster sample\label{sec:2}}

\subsection{Assumptions}

We assume that the cluster population follows scaling relations
derived from the spherical collapse model. The virial radius $r_{200}$
is defined such that the mean overdensity of the halo within $r_{200}$
is 200 times the critical density. This implies the relation
\begin{equation}
  r_{200}=\left(\frac{GM}{100\,H^2(z)}\right)^{1/3}
\label{eq:0.0}
\end{equation}
between $r_{200}$ and halo mass $M$. The Hubble function at redshift
$z$ is $H(z)$. Temperature $T$ and virial mass $M$ are assumed to be
related by
\begin{equation}
  kT=(kT)_{15}\,M_{15}^{2/3}\,(1+z)\,
  \left(\frac{\Omega_0}{\Omega(z)}\right)^{1/3}\,
  \left(\frac{\Delta_\mathrm{c}}{178}\right)^{1/3}
\label{eq:0.1}
\end{equation}
(e.g.~Eke et al. 1996; Navarro et al. 1995), where
$M_{15}=M/10^{15}\,M_\odot$ and $(kT)_{15}=6.03\,\mathrm{keV}$ is the
temperature of a cluster with $M_{15}=1$ (Mathiesen \& Evrard 2000).
The density parameter at redshift $z$ is $\Omega(z)$, and
$\Delta_\mathrm{c}$ is the mean overdensity of a virialised sphere,
\begin{equation}
  \Delta_\mathrm{c}=9\pi^2\,
  \left[1+\alpha\,(\Omega-1)+\Omega^\beta\right]
\label{eq:0.2}
\end{equation}
with
\begin{equation}
  (\alpha,\beta)=\left\{\begin{array}{ll}
    (0.1210,0.6756) & \hbox{open} \\
    (0.7076,0.4403) & \hbox{flat} \\
  \end{array}\right\}\;\hbox{cosmology}
\label{eq:0.3}
\end{equation}
(Stoehr 1999).

Moreover, we assume that the total number $N$ of thermal electrons
within a cluster's virial radius is proportional to the virial mass,
\begin{equation}
  N_\mathrm{e}=\frac{1+f_\mathrm{H}}{2}\,f_\mathrm{B}\,
  \frac{M}{m_\mathrm{p}}\;,
\label{eq:0.4}
\end{equation}
where $f_\mathrm{B}$ is the baryon fraction of the cluster mass,
$f_\mathrm{H}$ is the hydrogen fraction of the baryonic mass,
$f_\mathrm{H}\approx0.76$, and $m_\mathrm{p}$ is the proton mass.
From X-ray data of an ensemble of 45 clusters, Mohr et al.~(1999)
derived $f_\mathrm{B}=0.075\,h^{-3/2}$. Myers et al.~(1997) used
thermal Sunyaev-Zel'dovich observations in three nearby Abell clusters
to find a lower baryonic mass fraction of
$f_\mathrm{B}=0.061\,h^{-1}$, but this interpretation strongly depends
on cluster shape (cf.~Grego et al.~2000). We adopt the baryon fraction
by Mohr et al.~below.

We describe the projected thermal electron density with a King
profile, i.e.~a $\beta$ model with $\beta=1$,
\begin{equation}
  n_\mathrm{e,2D}(\theta)=n_0\,\left[1+\left(
  \frac{\theta}{\theta_\mathrm{c}}\right)^2\right]^{-1}\;,
\label{eq:0.4a}
\end{equation}
where $\theta_\mathrm{c}$ is the angular core radius. For our choice
of $\beta=1$, the entropy-driven cluster evolution model (Bower 1997)
predicts
\begin{equation}
  r_\mathrm{c}=r_\mathrm{c0}\,(1+z)^{-1/2(1+\epsilon)}\;,
\label{eq:0.4b}
\end{equation}
where $\epsilon$ is the entropy parameter. We put $\epsilon=0$ here
corresponding to the constant-entropy model (Kaiser 1991; Evrard \&
Henry 1991), resulting in a much more gentle redshift evolution of the
core radius than predicted by the self-similar evolution model (Kaiser
1986). However, for poorly resolved cluster observations like those
expected from {\em Planck\/}, the choice of $\epsilon$ only marginally
affects the results. We further choose
$r_\mathrm{c0}=0.13\,\hbox{Mpc}/h$.

Conventionally, the number density of dark-matter haloes is described
by the Press-Schechter model (Press \& Schechter 1974). The
Press-Schechter mass function can be written as
\begin{eqnarray}
  n_\mathrm{PS}(M,z)&=&\frac{\bar\rho}{\sqrt{2\pi}\,D_+(z)\,M^2}\,
  \left(1+\frac{n}{3}\right)\,
  \left(\frac{M}{M_\ast}\right)^{(n+3)/6}\nonumber\\
  &\times&\exp\left[-\frac{1}{2\,D_+^2(z)}\,
  \left(\frac{M}{M_\ast}\right)^{(n+3)/3}\right]\;,
\label{eq:0.5}
\end{eqnarray}
where $M_\ast$ and $\bar\rho$ are the nonlinear mass today and the
mean background density {\em at the present epoch\/}, and $D_+(z)$ is
the linear growth factor of density perturbations, normalised to unity
today, $D_+(0)=1$. Finally, $n$ is the effective exponent of the
dark-matter power spectrum at the cluster scale, $n\approx-1$.

Sheth \& Tormen (1999) recently modified the mass function
(\ref{eq:0.5}), and Sheth et al.~(1999) introduced ellipsoidal rather
than spherical collapse. Jenkins et al.~(1999) derived the mass
function of dark-matter haloes from numerical simulations and found a
fitting formula very close to Sheth \& Tormen's, but with lower
amplitude at the high-mass end. We used all three mass functions here
and found that deviations from the Press-Schechter prediction
noticeably change the results, so we used the fitting formula by
Jenkins et al.~for the results shown.

\subsection{Cluster detection}

The Sunyaev-Zel'dovich effect is determined by the Compton-$y$
parameter,
\begin{equation}
  y(\vec\theta)=\frac{kT}{m_\mathrm{e}c^2}\sigma_\mathrm{T}\,
  \int\d l\,n_\mathrm{e}(\vec\theta,l)\;,
\label{eq:2.1}
\end{equation}
if the gas distribution in the cluster is isothermal. The thermal,
three-dimensional electron density is written as
$n_\mathrm{e}(\vec\theta,l)$ to indicate that it depends on the
direction $\vec\theta$ on the sky, and $\sigma_\mathrm{T}$ is the
Thomson scattering cross section.

In the absence of background noise, the total Compton-$y$ parameter
seen from a galaxy cluster would be
\begin{equation}
  Y=\int\d^2\vec\theta\,y(\vec\theta)=
  \frac{kT}{m_\mathrm{e}c^2}\,
  \frac{\sigma_\mathrm{T}}{D_\mathrm{d}^2}\,N_\mathrm{e}\;,
\label{eq:2.2}
\end{equation}
where $D_\mathrm{d}$ is the angular-diameter distance to the cluster
and $N_\mathrm{e}$ is the total number of (thermal) electrons in the
cluster; see Eq.~(\ref{eq:0.4}). Since $y$ is dimension-less, $Y$ is
effectively a solid angle.

The angular resolution of {\em Planck\/} will not allow to spatially
resolve low-mass clusters, and even high-mass clusters will be barely
resolved (Aghanim et al. 1997; Hobson et al. 1998). There will
therefore be a Compton-$y$ background $y_\mathrm{bg}$ dominated by
low-mass clusters, since their much higher number density
over-compensates their lower individual contributions. An ideally
isotropic background would not matter since it could be completely
removed. It is therefore the average background fluctuation level,
$\Delta y_\mathrm{bg}$, that we have to take into account. Moreover,
{\em Planck\/} will see the clusters convolved with its beam profile
$b(\vec\theta)$.

We therefore adopt the following cluster detection criterion for {\em
Planck\/} (see also Bartlett 2000). Let the beam-convolved Compton-$y$
profile of a cluster be
\begin{equation}
  \bar y(\vec\theta)=\int\d^2\theta'\,y(\vec\theta')\,
  b(\vec\theta-\vec\theta')\;.
\label{eq:2.3}
\end{equation}
A cluster is assumed to be detectable by {\em Planck\/} if its
integrated, beam-convolved Compton-$y$ parameter is sufficiently
large, i.e.
\begin{equation}
  \bar Y=\int\d^2\theta\,\bar y(\vec\theta)
  \ge\bar Y_\mathrm{min}\;,
\label{eq:2.4}
\end{equation}
where the integral covers the area where the integrand sufficiently
exceeds the background fluctuations, $\bar y\ge\nu\Delta
y_\mathrm{bg}$. We choose $\nu=3$ in the following.

In evaluating (\ref{eq:2.4}), we use assumption (\ref{eq:0.4a}) that
the thermal electron density follows a King profile. Then,
\begin{equation}
  y(\vec\theta)=y_0\,\left[1+\left(
    \frac{\theta}{\theta_\mathrm{c}}
  \right)^2\right]^{-1}\;,
\label{eq:2.5}
\end{equation}
with the angular core radius $\theta_\mathrm{c}$. Approximating the
beam profile with a Gaussian,
\begin{equation}
  b(\vec\theta)=\frac{1}{2\pi\sigma_\mathrm{B}}\,
  \exp\left(-\frac{\vec\theta^2}{2\sigma_\mathrm{B}^2}\right)\;,
\label{eq:2.6}
\end{equation}
the beam-convolved Compton-$y$ profile is,
\begin{equation}
  \bar y(\vec\theta)=2a\mathrm{e}^{-ax^2}\,
  \int_0^\infty\,x'\d x'\,
  \frac{\exp(-ax^{\prime\,2})}
  {1+x^{\prime\,2}}\mathrm{I}_0(2axx')\;,
\label{eq:2.7}
\end{equation}
where $a\equiv\theta_\mathrm{c}^2/2\sigma_\mathrm{B}^2$,
$x\equiv\theta/\theta_\mathrm{c}$, and $\mathrm{I}_0(x)$ is the
zeroth-order modified Bessel function of the first kind. For wide
beams, as in the case of {\em Planck\/}, the integral (\ref{eq:2.7}) can be
well approximated by
\begin{eqnarray}
  \bar y(\theta)&\approx&\frac{a\mathrm{e}^{-ax^2}}{4}\,
  \left[
    ax^2(4+ax^2(1-a))\right.\nonumber\\
  &+&\left.
    \mathrm{e}^a(a^2x^2-2)^2\mathrm{E}_1(a)
  \right]\;,
\label{eq:2.8}
\end{eqnarray}
where $\mathrm{E}_1(a)$ is the first-order exponential integral. This
expression has the advantage of being much more easily evaluated
numerically than the integral (\ref{eq:2.7}). Ignoring the background
fluctuations, $\bar Y$ from (\ref{eq:2.4}) equals $Y$ from
(\ref{eq:2.2}), as it should.

Finally, the background level is given by
\begin{eqnarray}
  y_\mathrm{bg}&=&\int\d z\,\left|\frac{\d V}{\d z}\right|\,
  (1+z)^3\int\d M\,n_\mathrm{PS}(M,z)\,Y(M,z)\nonumber\\
  &=& \int\d M\int\d V\,Y(M,z)\,\frac{\d^2N(M,z)}{\d M\d V}\;,
\label{eq:2.9}
\end{eqnarray}
where $\d V$ is the cosmic volume per unit redshift and unit solid
angle, $n_\mathrm{PS}(M,z)$ is the cluster mass function
(\ref{eq:0.5}), and $Y(M,z)$ is the integrated Compton-$y$ parameter
from (\ref{eq:2.2}) expressed in terms of cluster mass $M$ and
redshift $z$ (e.g.~Barbosa et al. 1996; Da Silva et
al. 1999). Neglecting cluster correlations, background fluctuations
are due to Poisson fluctuations in the number of clusters per unit
mass and volume. The {\em rms\/} background fluctuation is then
\begin{equation}
  \Delta y_\mathrm{bg}=\left[\int\d M\int\d V\,Y^2(M,z)\,
  \frac{\d^2N(M,z)}{\d M\d V}\right]^{1/2}\;.
\label{eq:2.9a}
\end{equation}
Although only undetected clusters contribute to the background and its
fluctuation, the mass integral can be extended to infinity because of
the steep decline of the mass function. While $y_\mathrm{bg}$ is a few
times $10^{-6}$ depending on cosmology (e.g.~Barbosa et al.~1996; da
Silva et al.~2000), the background fluctuation $\Delta y_\mathrm{bg}$
increases from $5\times10^{-8}$ for $\Lambda$CDM to $1.3\times10^{-7}$
for $\Omega_0=1$.

The sensitivity limit $Y_\mathrm{min}$ can be derived from the nominal
(antenna) temperature sensititvity of {\em Planck\/}, i.e.~$\Delta
T/T\ge2\times10^{-6}$ (Puget et al. 1998). The beam-integrated
Sunyaev-Zel'dovich effect changes the flux by $\Delta
F_\nu=2\,Y\,I_\nu$. This leads to a change $\Delta I_\nu=\Delta
F_\nu/\delta\Omega$ in specific intensity across a beam of solid angle
$\delta\Omega$. The two expressions imply
\begin{equation}
  Y\ge\frac{\Delta I_\nu}{I_\nu}\,\frac{\delta\Omega}{2}=
  \frac{\Delta T}{T}\,\frac{\delta\Omega}{2}\;.
\label{eq:2.10}
\end{equation}
With $\delta\Omega\approx50\,\mathrm{arcmin}^2$ for the relevant {\em
Planck\/} beams, expression (\ref{eq:2.10}) implies
$Y_\mathrm{min}\approx10^{-4}\,\mathrm{arcmin}^2$ (see also Haehnelt
1997). Unless mentioned otherwise, we will assume
$Y_\mathrm{min}=3\times10^{-4}\,\mathrm{arcmin}^2$ below for a
conservative limit.

\begin{figure}[ht]
  \centerline{\includegraphics[width=\fsize]{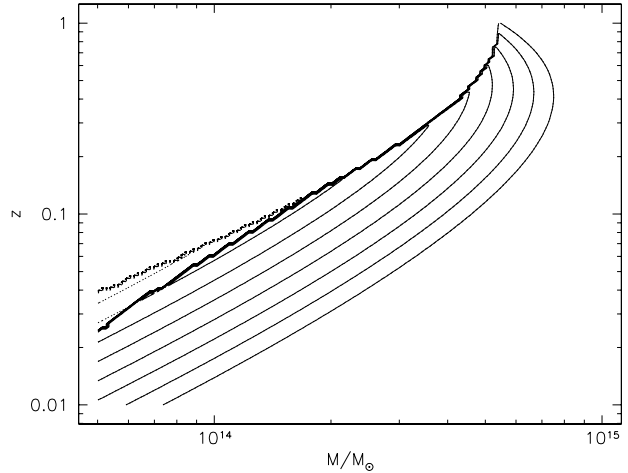}}
\caption{Contours of the number density in the $M$-$z$ plane of
clusters which will be detectable for {\em Planck\/}. The lowest
contour level is at $10^{-10.5}\,M_\odot^{-1}$, and the levels are
logarithmically spaced at $0.2$ dex. Dotted contours ignore the
effects of beam convolution and background fluctuations. Note the
logarithmic axes.}
\label{fig:1}
\end{figure}

Figure~\ref{fig:1} shows contours of the number density in the $M$-$z$
plane of clusters detectable by {\em Planck\/}. A low-density, flat
CDM universe with $\Omega_0=0.3$, $\Omega_\Lambda=0.7$ and $h=0.7$ was
adopted and normalised to match the local abundance of massive
clusters (White et al. 1993; Eke et al. 1996, Viana \& Liddle
1996). Following the results by Mohr et al.~(1999), the baryon
fraction is set to $f_\mathrm{B}=13\%$. The contours start at
$10^{-10.5}\,M_\odot^{-1}$ and are separated by $0.2$~dex. While the
solid contours were obtained using the detection criteria discussed
above, the dotted contours ignore the effects of beam convolution and
background contamination. The axes are plotted logarithmically to
emphasise the effect of the detailed detection criteria.

\begin{figure}[ht]
  \centerline{\includegraphics[width=\fsize]{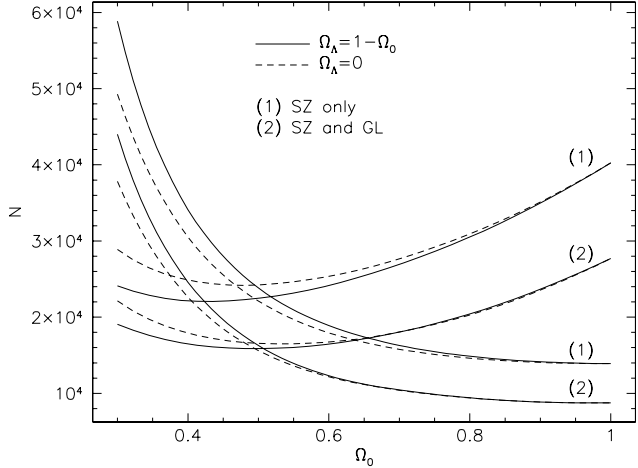}}
\caption{The number of detectable Sunyaev-Zel'dovich clusters is
plotted against $\Omega_0$, assuming $\Omega_\Lambda=0$ or
$\Omega_\Lambda=1-\Omega_0$. Four pairs of curves are plotted. The
solid curve of each pair was calculated for a flat, the dashed curve
for an open model with given $\Omega_0$. The almost horizontal curves
assume $f_\mathrm{B}=0.13$. For the monotonically falling set of
curves, $f_\mathrm{B}$ varies according to the primordial
nucleosynthesis constraint. The upper curves ignore lensing, the lower
curves take it into account, as indicated.}
\label{fig:2}
\end{figure}

Taking beam convolution and background into account, the high-redshift
part of the low-mass Sunyaev-Zel'dovich cluster population is
removed. The background fluctuations impede in several ways on the
possibility to discriminate cosmologies through the number of
Sunyaev-Zel'dovich clusters detectable for {\em Planck\/}, as da Silva
et al.~(2000) suspected earlier. On the one hand, reducing $\Omega_0$
leads to slower evolution of the cluster number density (Richstone et
al. 1992) and larger cosmic volume per unit redshift, but a
substantial part of these clusters is not visible to {\em Planck\/}
because of the dependence of the integrated $y$ parameter on
distance. The background increases with cluster number, so the
relative background fluctuations decrease. The net effect is a very
minor change of the detectable cluster number with cosmology, as shown
in Fig.~\ref{fig:2}.

The figure shows four pairs of curves. One curve per pair represents a
flat universe, the other an open universe, as indicated. The almost
horizontal curves were calculated for a constant baryon fraction,
$f_\mathrm{B}=0.13$, while the baryon fraction varies according to the
constraint from primordial nucleosynthesis,
$f_\mathrm{B}=\Omega_\mathrm{B}/\Omega_0=0.024/\Omega_0\,h^2$
(e.g.~Schramm 1998 for a review). The upper curves show the number of
Sunyaev-Zel'dovich clusters detectable for {\em Planck\/}. The lower
curves take gravitational lensing into account, as detailed in
Sect.~\ref{sec:3} below. For constant baryon fraction, the cluster
number changes with $\Omega_0$ by at most a factor of two, and it
drops by a factor of $\sim3$ if the baryon fraction is fixed by
primordial nucleosynthesis. The gravitationally lensing subsample
comprises more than $70\%$ of the full sample, quite independent of
cosmology.

\section{Lensing effects of the cluster sample\label{sec:3}}

We give only a very brief summary of weak lensing by clusters
here. For a detailed review, see (Bartelmann \& Schneider 2000).

\subsection{Relevant properties of the Aperture Mass}

Schneider (1996) suggested quantifying the weak-lensing effects of
dark-matter haloes with the {\em aperture mass\/},
\begin{equation}
  M_\mathrm{ap}(\theta)=\int\d^2\vec\vartheta\,\kappa(\vec\vartheta)\,
  U(|\vec\vartheta|)\;,
\label{eq:3.1}
\end{equation}
which is an integral over the lensing convergence $\kappa$ within a
circular aperture with (angular) radius $\theta$, weighted by a
function $U(\vartheta)$ which vanishes outside the aperture. The prime
advantage of $M_\mathrm{ap}$ is that it can directly be determined
from the measured tidal distortions of background-galaxy images in the
chosen aperture, provided $U(\vartheta)$ is compensated, i.e.
\begin{equation}
  \int_0^\theta\d^2\vartheta\,\vartheta U(\vartheta)=0\;.
\label{eq:3.2}
\end{equation}
A broad class of weight functions satisfies this condition. We will
follow Schneider's suggestion and take
\begin{equation}
  U(\vartheta)=\frac{9}{\pi\theta^2}\,
  \left[1-\left(\frac{\vartheta}{\theta}\right)^2\right]
  \left[\frac{1}{3}-\left(\frac{\vartheta}{\theta}\right)^2\right]
\label{eq:3.3}
\end{equation}
within the aperture, and $U(\vartheta)=0$ outside.

Schneider (1996) also calculated the dispersion of $M_\mathrm{ap}$ due
to the finite number of randomly distributed background galaxies and
their intrinsic ellipticities. Assuming typical values for the number
density $n_\mathrm{g}$ of suitably bright background galaxies and for
the dispersion $\sigma_\epsilon$ of their intrinsic ellipticities, he
found
\begin{eqnarray}
  \sigma_\mathrm{M}(\theta)&=&0.016\,
  \left(\frac{n_\mathrm{g}}{30\,\mathrm{arcmin}^2}\right)^{-1/2}\,
  \left(\frac{\sigma_\epsilon}{0.2}\right)\nonumber\\
  &\times&
  \left(\frac{\theta}{1\,\mathrm{arcmin}}\right)^{-1}\;.
\label{eq:3.3a}
\end{eqnarray}

For a singular isothermal sphere with Einstein radius
$\theta_\mathrm{E}$, $\kappa(\vartheta)=\theta_\mathrm{E}/2\vartheta$,
and
\begin{equation}
  M_\mathrm{ap}^\mathrm{(SIS)}(\theta)=
  \frac{4}{5}\frac{\theta_\mathrm{E}}{\theta}\;.
\label{eq:3.4}
\end{equation}
The Einstein radius is proportional to the squared velocity dispersion
$\sigma_v^2$ of the singular isothermal sphere,
$\theta_\mathrm{E}\propto\sigma_v^2$. Since the virial mass scales
with velocity dispersion as $M\propto\sigma_v^{3}$, the aperture mass
essentially measures $M^{2/3}$ when applied to a singular isothermal
sphere. According to (\ref{eq:3.3a}), the signal-to-noise ratio of
$M_\mathrm{ap}$ for a singular isothermal sphere is independent of the
aperture size $\theta$, since both $M_\mathrm{ap}$ and
$\sigma_\mathrm{M}$ scale as $\theta^{-1}$ in this case.

The situation is somewhat more complicated for the density profile
suggested by Navarro, Frenk \& White (1997),
\begin{equation}
  \rho(r)=\frac{\rho_\mathrm{crit}\,\delta_\mathrm{c}}
  {(r/r_\mathrm{s})(1+r/r_\mathrm{s})^2}\;,
\label{eq:3.5}
\end{equation}
where $\rho_\mathrm{crit}$ is the critical density and
$\delta_\mathrm{c}$ is a characteristic overdensity. The
characteristic radial scale $r_\mathrm{s}$ is related to the virial
radius through $r_\mathrm{s}=r_{200}/c$, where $c$ is the
concentration parameter. Bartelmann (1996) showed that the NFW profile
has the lensing convergence
\begin{equation}
  \kappa(x)=\frac{2\kappa_\mathrm{s}}{1-x^2}\,\left(
    1-\frac{2}{\sqrt{1-x^2}}\mathrm{arctanh}\sqrt{\frac{1-x}{1+x}}
  \right)\;,
\label{eq:3.6}
\end{equation}
with the convergence scale
\begin{equation}
  \kappa_\mathrm{s}\equiv
  \frac{\rho_\mathrm{crit}\delta_\mathrm{c}r_\mathrm{s}}
  {\Sigma_\mathrm{cr}}\;,
\label{eq:3.7}
\end{equation}
where
\begin{equation}
  \Sigma_\mathrm{cr}=\frac{c^2}{4\pi G}\,
  \left(\frac{D_\mathrm{s}}{D_\mathrm{d}D_\mathrm{ds}}\right)
\label{eq:3.8}
\end{equation}
is the critical surface mass density of a gravitational lens. The
angular diameter distances to the lens, the source, and from the lens
to the source are $D_\mathrm{d,s,ds}$, respectively.

Navarro et al. (1997) described how the parameters $\delta_\mathrm{c}$
and $r_\mathrm{s}$ are related to the virial mass $M$ of the
halo. Hence, despite the two formal parameters in the profile
(\ref{eq:3.5}), it is entirely determined once the halo mass is
fixed. The statistics of dark-matter haloes described bt the NFW
density profile has also been investigated by Kruse \& Schneider
(1999).

Unfortunately, there is no closed expression for the aperture mass of
an NFW profile. Progress can be made, however, under the assumption
that the aperture radius $\theta$ is smaller than the angular scale
radius, $\theta<\theta_\mathrm{s}=r_\mathrm{s}/D_\mathrm{d}$. With
\begin{equation}
  t\equiv\frac{\theta}{\theta_\mathrm{s}}\;,
\label{eq:3.9}
\end{equation}
one can show by means of Taylor expansions that
\begin{eqnarray}
  M_\mathrm{ap}^\mathrm{(NFW)}(t)&\approx&\kappa_\mathrm{s}\,\left[
    1+t^2\left(\frac{19}{32}+\frac{3}{4}\ln\frac{t}{2}\right)
  \right.\nonumber\\
   &+&\left.
    t^4\left(\frac{77}{160}+\frac{3}{4}\ln\frac{t}{2}\right)
  \right]\;.
\label{eq:3.10}
\end{eqnarray}
In other words, $M_\mathrm{ap}^\mathrm{(NFW)}\to\kappa_\mathrm{s}$ in
the limit of small apertures. Within the mass range of galaxy
clusters, $\rho_\mathrm{crit}\delta_\mathrm{c}$ is essentially a
constant. Given the geometry of the lens system,
$M_\mathrm{ap}^\mathrm{(NFW)}$ therefore essentially measures the
scale radius $r_\mathrm{s}$ or, equivalently, $M^{1/3}$.

\begin{figure}[ht]
  \centerline{\includegraphics[width=\fsize]{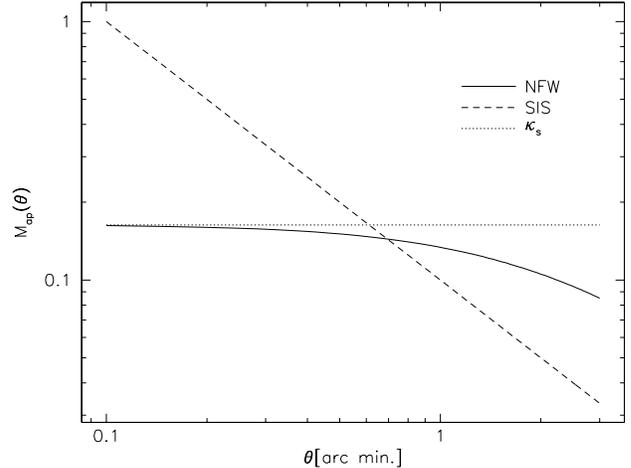}}
\caption{The aperture mass $M_\mathrm{ap}(\theta)$ is plotted against
aperture radius $\theta$ for an NFW halo of mass
$5\times10^{14}\,M_\odot$ at redshift $0.2$ lensing sources at
redshift $1.5$. The dotted curve shows $\kappa_\mathrm{s}$, the dashed
curve illustrates the $1/\theta$ behaviour of $M_\mathrm{ap}(\theta)$
for a singular isothermal sphere. As explained in the text,
$M_\mathrm{ap}\to\kappa_\mathrm{s}$ for $\theta\to0$, and
$M_\mathrm{ap}$ depends much more weakly on aperture size for an NFW
halo than for a SIS.}
\label{fig:3a}
\end{figure}

\begin{figure}[ht]
  \centerline{\includegraphics[width=\fsize]{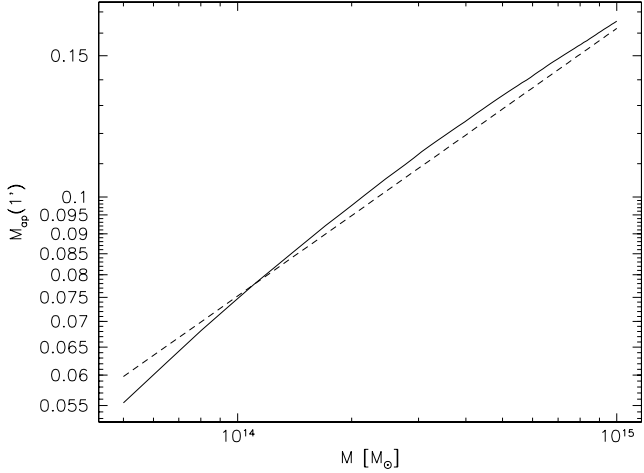}}
\caption{The aperture mass $M_\mathrm{ap}(\theta)$ with $\theta=1'$ is
plotted against virial mass for haloes at redshift $0.2$ lensing
sources at redshift $1.5$. The dashed curve shows the
$M_\mathrm{ap}\propto M^{1/3}$ behaviour expected for sufficiently
massive haloes.}
\label{fig:3b}
\end{figure}

We thus see two important differences between the aperture masses of
singular isothermal spheres and NFW profiles: For the latter, the
aperture mass depends much more weakly on the halo mass and on the
aperture size than for the former. Among other things, this implies
with Eq.~(\ref{eq:3.3a}) that the signal-to-noise ratio for the
aperture mass of an NFW halo decreases with decreasing aperture
size. Figures~\ref{fig:3a} and \ref{fig:3b} illustrate these results.

\subsection{Dependence on source redshifts}

Clearly, $M_\mathrm{ap}$ depends on the redshift of the background
sources used to measure the gravitational shear. In presence of a
(normalised) source redshift distribution $p(z_\mathrm{s})$, the
aperture mass of a halo at redshift $z$ becomes
\begin{equation}
  \bar M_\mathrm{ap}(\theta;z)=\int_z^\infty\d z_\mathrm{s}\,
  M_\mathrm{ap}(\theta;z,z_\mathrm{s})\,p(z_\mathrm{s})\;.
\label{eq:3.11}
\end{equation}
A useful and sufficiently accurate representation of the source
redshift distribution is
\begin{equation}
  p(z_\mathrm{s})=\frac{\beta}{z_0^3\Gamma(3/\beta)}\,
  z^2\,\exp\left[-\left(\frac{z}{z_0}\right)^\beta\right]\;,
\label{eq:3.12}
\end{equation}
for which we choose $z_0=1.0$ and $\beta=1.5$ (Smail et
al. 1995). Since $M_\mathrm{ap}$ is linear in $\kappa$,
(\ref{eq:3.11}) amounts to replacing $\kappa_\mathrm{s}$ in
(\ref{eq:3.6}) by its source-redshift averaged counterpart. Unless
stated otherwise, the source redshift average (\ref{eq:3.11}) will
from now on implicitly be applied.

\subsection{Lensing Sunyaev-Zel'dovich clusters}

We can now ask: What fraction of the cluster sample detectable for
{\em Planck\/} will produce significant lensing effects? The
significance of weak lensing by any given cluster of mass $M$ and
redshift $z$ can be estimated by means of the signal-to-noise ratio
$\mathcal{S}(\theta)=M_\mathrm{ap}(\theta)/\sigma_\mathrm{M}(\theta)$,
where $\sigma_\mathrm{M}(\theta)$ is the dispersion
(\ref{eq:3.3a}). Then, the condition
$\mathcal{S}\ge\mathcal{S}_\mathrm{min}$ defines a region in the
$M$-$z$ plane within which clusters produce a significant weak-lensing
effect. We note that the numerical evaluation of that condition needs
to take into account that the dispersion $\sigma_\mathrm{M}(\theta)$
depends on cluster redshift because the number density of sources,
$n_\mathrm{g}$, does: Only sources at redshifts higher than the
cluster contribute to the measured tidal field.

The sample of clusters which can be detected by {\em Planck\/} {\em
and\/} optically through their lensing effect is then determined by
the two conditions
\begin{equation}
  Y\ge Y_\mathrm{min}\quad\hbox{and}\quad
  \mathcal{S}\ge\mathcal{S}_\mathrm{min}\;.
\label{eq:3.13}
\end{equation}
We will assume $\mathcal{S}_\mathrm{min}=5$ in the following.

\begin{figure}[ht]
  \centerline{\includegraphics[width=\fsize]{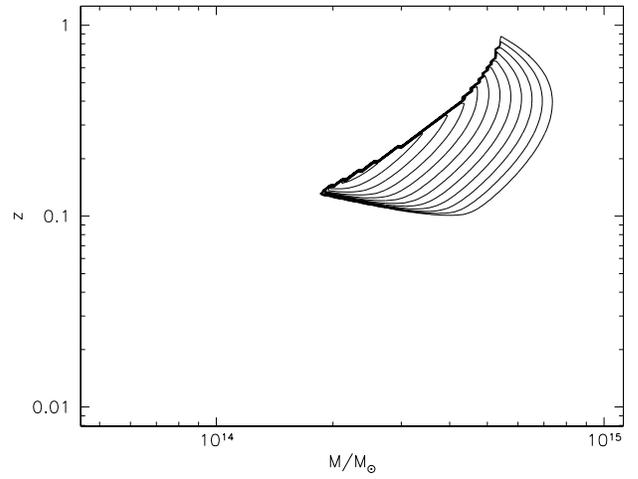}}
\caption{Contours of the number density in the $M$-$z$ plane of
Sunyaev-Zel'dovich clusters detectable by {\em Planck\/} that are also
efficient weak lenses. The contours range within
$(0.1-3.0)\times10^{-10}\,M_\odot^{-1}$ in steps of
$0.2\times10^{-10}\,M_\odot^{-1}$. A flat, low-density,
cluster-normalised CDM model was assumed.}
\label{fig:4}
\end{figure}

Figure~\ref{fig:4} shows contours of the number density in the $M$-$z$
plane of Sunyaev-Zel'dovich clusters that are efficient weak
lenses. Although the sample appears to be substantially reduced,
Fig.~\ref{fig:2} shows that this is not the case: The efficiently
lensing Sunyaev-Zel'dovich cluster sample still comprises more than
$70\%$ of the original Sunyaev-Zel'dovich sample. In other words, the
majority of clusters that {\em Planck\/} will detect as
Sunyaev-Zel'dovich sources will also be detectable in the optical as
efficient gravitational lenses. On the whole sky, a few times $10^4$
such clusters are expected, quite independent of cosmological
parameters (see Fig.~\ref{fig:2}; cf.~Barbosa et al. 1996; Da Silva et
al. 2000; Hern\'andez-Monteagudo et al. 2000).

\begin{figure}[ht]
  \centerline{\includegraphics[width=\fsize]{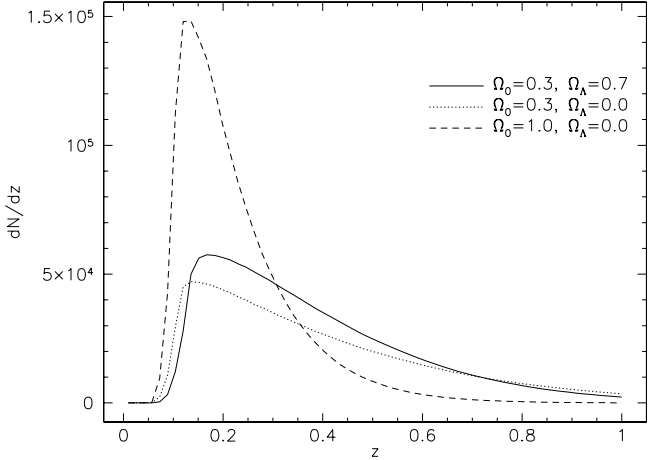}}
\caption{Redshift distributions of lensing Sunyaev-Zel'dovich clusters
for three different cosmological models, as indicated. The redshift
distribution for $\Omega_0=1$ is cut off by rapid cluster
evolution. For all cosmological models, the majority of clusters sits
near $z=0.1-0.2$.}
\label{fig:5}
\end{figure}

Combining the thermal Sunyaev-Zel'dovich effect and significant weak
gravitational lensing therefore defines a large cluster
sample. Figure~\ref{fig:4} illustrates that a large portion of this
sample is remarkably confined in redshift. Because of the exponential
fall-off of the cluster population towards high mass, the sharp
cut-off of the sample towards high redshift due to the
Sunyaev-Zel'dovich selection criterion, and the fairly steep decline
of lensing efficiency towards low redshifts, the majority of the
combined Sunyaev-Zel'dovich-lensing cluster sample is confined to a
narrow redshift interval around $z\approx0.2$. This is further
illustrated by the redshift distributions in Fig.~\ref{fig:5}.

\subsection{Analytic Description of the Sample}

The main features of the region in $M$-$z$ space occupied by lensing
Sunyaev-Zel'dovich clusters can approximately be described
analytically. Evidently, there are two sharp edges in the $M$-$z$
plane which truncate the cluster population. The upper and lower
edges, respectively, are due to the Sunyaev-Zel'dovich and lensing
detection criteria. At low masses and redshifts, both edges are almost
straight lines in the double-logarithmically plotted Fig.~\ref{fig:4},
i.e.~they are power laws. At $M\approx5\times10^{14}\,M_\odot$, the
upper edge quite abruptly turns into the vertical. We will now explain
these features.

Ignoring beam convolution and background effects, the upper edge
approximately follows the line $z\propto M^{5/6}$: The integrated
Compton-$y$ parameter scales like $Y\propto TM/D_\mathrm{d}^2$, and
the temperature $T\propto M^{2/3}$. For low redshifts,
$D_\mathrm{d}\propto z$, hence $Y\propto M^{5/3}/z^2$. The relation
$z\propto M^{5/6}$ then follows from the definition of the detection
edge by $Y=Y_\mathrm{min}=\mathrm{constant}$. Figure~\ref{fig:1} shows
that the sample obtained with realistic detection criteria shows the
same behaviour for cluster masses above $10^{14}\,M_\odot$.

An interesting deviation from this relation occurs at higher mass and
redshift (see also Holder et al.~1999). The cluster temperature at a
given mass increases with redshift like $(1+z)$, and the integrated
Compton-$y$ parameter diminishes with distance like
$D_\mathrm{d}^{-2}$. The angular-diameter distance $D_\mathrm{d}$
rises $\propto z$ for small redshifts and then flattens off. At a
certain redshift, call it $z_1$, the ratio $D_\mathrm{d}^2(z)/(1+z)$
reaches a maximum, so that the integrated Compton-$y$ parameter for a
cluster of fixed mass first decreases with redshift out to $z_1$, and
then increases again. For an Einstein-de Sitter universe, this happens
for $z>z_1=7/9$, and for slightly higher $z_1$ in lower-density
universes. Let $M_1$ be the lower mass limit for a cluster to be seen
out to redshift $z_1$, then {\em all\/} clusters with masses $M\ge
M_1$ will be detectable as Sunyaev-Zel'dovich sources, and they will
also be significant lenses unless the background source distribution
is narrow and located at low redshifts. Recent detections of
efficienly lensing clusters at high redshifts (Luppino \& Kaiser 1997,
Clowe et al.~2000) show that this is unlikely to be the case. This
explains the upturn of the upper edge of the contours in
Fig.~\ref{fig:4} near $M\approx5\times10^{14}\,M_\odot/h$. We should
note, however, that the given explanation rests upon the assumption
that clusters of fixed mass are hotter at higher redshifts, and that
$T\propto(1+z)$. Should the true relation between temperature and
redshift be flatter or steeper, the lower mass limit $M_1$ would
increase or decrease, respectively. The sudden extension of the sample
for $M\ge M_1$ therefore contains indirect information on the thermal
history of the clusters. We will return to this point later.

On the other hand, the lower edge approximately follows $z\propto
M^{-1/3}$: For low-redshift clusters, the ratio between the distances
from the lens to the source and from the observer to the source is
almost unity, hence the critical surface mass density is proportional
to $D_\mathrm{d}^{-1}$ only. For low redshifts, $D_\mathrm{d}\propto
z$, so that $\kappa_\mathrm{s}\propto z$. The lensing criterion
$\mathcal{S}\ge\mathcal{S}_\mathrm{min}$ therefore implies $z
r_\mathrm{s}\propto zM^{1/3}\ge\hbox{const}$.

\section{Mass distribution of the combined sample\label{sec:4}}

\subsection{Characteristic shape}

\begin{figure}[ht]
  \centerline{\includegraphics[width=\fsize]{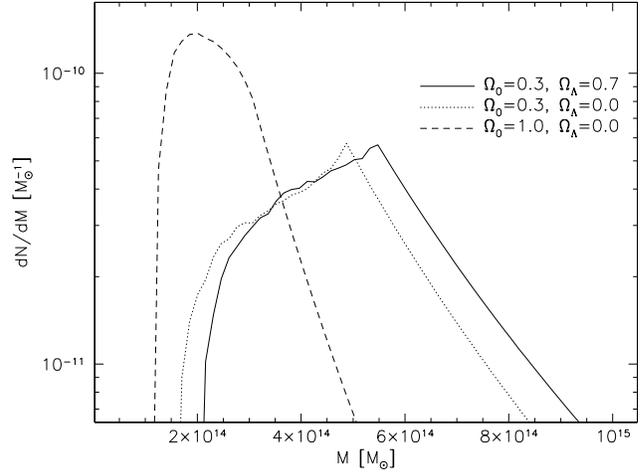}}
\caption{Mass distribution of lensing Sunyaev-Zel'dovich clusters for
three different cosmologies, as indicated. For low cosmic density, the
curves have a characteristic shape: They rise abruptly near
$2\times10^{14}\,M_\odot$, flatten, reach a peak near
$5\times10^{14}\,M_\odot$, and then fall off exponentially. The mass
distribution for high $\Omega_0$ is shifted to lower masses and lacks
the peak.}
\label{fig:6}
\end{figure}

Figure~\ref{fig:6} shows that the mass distribution of lensing
Sunyaev-Zel'dovich clusters has a characteristic shape, at least for
low-density cosmologies.

The whole cluster sample is confined to the fairly narrow mass
range. For low $\Omega_0$, $2\times10^{14}\le
M/M_\odot\le8\times10^{14}$, and even narrower for high
$\Omega_0$. For low cosmic density, the mass distribution function
abruptly rises at the low-mass end, flattens, peaks near
$5\times10^{14}\,M_\odot$, and then falls off exponentially. While the
peak is pronounced for the low-density cosmologies, it is absent when
$\Omega_0$ is high.

Using the approximate analytic sample descriptions derived earlier for
the number density of significantly lensing Sunyaev-Zel'dovich
clusters in the $M$-$z$ plane, the shape of the mass distribution is
straightforwardly explained. All clusters can be seen above the mass
$M_1$ introduced above. The high-mass end of the mass distribution
therefore reflects the exponential cut-off expected from the
Press-Schechter mass function (\ref{eq:0.5}). The peak in the mass
distribution is given by $M_1$, i.e.~the minimum mass for clusters to
be visible out to the redshift $z_1$ where the squared
angular-diameter distance, divided by $(1+z)$, reaches its maximum.

The cut-off at the low-mass end represents the lower-left corner of
the contours in Fig.~\ref{fig:4}. It is defined by the intersection of
two lines, $z=C_\mathrm{GL}\,M^{-1/3}$ from lensing, and
$z=C_\mathrm{SZ}\,M^{5/6}$ from the Sunyaev-Zel'dovich effect, as
explained before. We are dealing with low redshifts, so distances may
be approximated by Hubble's law. For the low-density cosmological
models, we can further neglect the redshift evolution of the cluster
sample, i.e.~we can set $D_+(z)=1$ in (\ref{eq:0.5}). The number
distribution of detectable clusters as a function of $M$ can then be
approximated by
\begin{eqnarray}
  \frac{\d N}{\d M}&\propto&
  \left(C_\mathrm{SZ}^3\,M^{5/6}-C_\mathrm{GL}^3\,M^{-8/3}\right)
  \nonumber\\&\times&
  \exp\left[-\frac{1}{2}\left(\frac{M}{M_\ast}\right)^{2/3}\right]\;,
\label{eq:3.14}
\end{eqnarray}
where we have inserted the Press-Schechter mass function
(\ref{eq:0.5}) and put $n=-1$. This qualitatively explains the sudden
rise in the mass distributions for the low-$\Omega_0$ universes. For
high $\Omega_0$, cluster evolution is substantially faster. Therefore,
there are essentially no clusters reaching the mass- and redshift
limits $(M_1,z_1)$ leading to the pronounced peak in the mass
distributions. The low-mass cutoff, however, remains determined by the
intersection of the Sunyaev-Zel'dovich and lensing sample edges.

\subsection{Baryon Fraction and Hubble Constant}

If observable, the distinct features in the mass distribution of
lensing Sunyaev-Zel'dovich clusters have immediate physical
implications. The first was already mentioned: All else fixed, the
peak in the mass distribution contains indirect information on the
thermal evolution of the cluster sample with redshift. Second, the
exponential cut-off at the high-mass end allows to determine the
cluster mass function at redshift $z\sim0.2$, where the majority of
the cluster sample is located. Third, the sharp cut-off at the
low-mass end measures the intersection point between the two lines
$z=C_\mathrm{GL}\,M^{-1/3}$ and $z=C_\mathrm{SZ}\,M^{5/6}$. For a
fixed cosmological model, $C_\mathrm{GL}$ is fixed, while
$C_\mathrm{SZ}$ is $\propto f_\mathrm{B}h$. Equating the two
expressions leads to
\begin{equation}
  M_0\propto(f_\mathrm{B}h)^{-6/7}\;.
\label{eq:3.15}
\end{equation}
The higher the baryon fraction is, the lower is the mass limit for
detectable Sunyaev-Zel'dovich clusters. Likewise, for higher Hubble
constant, the angular-diameter distance to a fixed redshift is
smaller, and lower-mass clusters can be seen. Third, the location of
the peak, if it exists, is determined by the lowest mass required for
a cluster to be seen beyond the maximum in $D^2(z)/(1+z)$. This is
fixed by the Sunyaev-Zel'dovich detection criterion alone. It then
follows from (\ref{eq:2.2}) that
\begin{equation}
  M_1\propto(f_\mathrm{B}h)^{-3/5}\;;
\label{eq:3.16}
\end{equation}
for increasing baryon fraction, the total cluster mass can be lower
for the cluster to be seen, and likewise for a higher Hubble
constant. Although in reality the relations between the low-mass
cut-off and the peak position on the one hand and the baryon fraction
on the other will be more complicated, these considerations show that
the distinct features in the mass distribution contain information on
various physical properties of the cluster population at moderate
redshifts.

\begin{figure}[ht]
  \centerline{\includegraphics[width=\fsize]{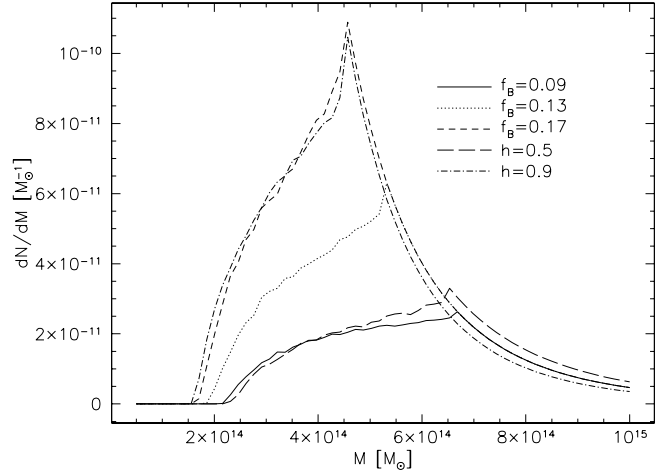}}
\caption{Mass distributions in the combined Sunyaev-Zel'dovich and
gravitational-lensing cluster sample, for different baryon fractions
$f_\mathrm{B}$ and fixed Hubble constant, and for different Hubble
constants and fixed baryon fraction, as indicated in the plot. The
solid, dotted, and short-dashed curves show that an increasing baryon
fraction shifts the peak to lower masses, as expected, and increases
the peak height, while keeping the exponential cut-off constant. The
dot-dashed, dotted, and long-dashed curves show that an increasing
Hubble constant also shifts the peak towards lower masses, but also
leads to lower amplitude in the exponential cut-off.}
\label{fig:7}
\end{figure}

It appears from the foregoing discussion of the location of cut-off
and peak in the mass function of the combined cluster sample, that
baryon fraction and Hubble constant shift them in a completely
parallel way. The real situation is somewhat more complex, for two
reasons: First, Hubble constant and baryon fraction are further
related through constraints from primordial nucleosynthesis. If
$\Omega_\mathrm{B}$ is the density parameter of baryons,
$f_\mathrm{B}=\Omega_\mathrm{B}/\Omega_0$, but primordial
nucleosynthesis determines the physical baryon density, hence
$\Omega_\mathrm{B}\,h^2$. Taking this into account, the baryon
fraction itself scales with the Hubble constant as
$f_\mathrm{B}\propto h^{-2}$. The diminution of the Sunyaev-Zel'dovich
effect with decreasing Hubble constant as a consequence of increasing
physical distances is therefore more than cancelled if the cluster
baryon fraction follows the primordial nucleosynthesis
constraint. Second, the Hubble constant also determines the shape
parameter of the dark-matter power spectrum,
$\Omega_0\,h$. Normalising the spectrum to the local abundance of rich
clusters keeps the spectrum fixed at the linear cluster scale of
$\approx10\,\hbox{Mpc}/h$. Lowering $h$ shifts the peak of the
spectrum to larger scales, leading to more power on larger and less
power on smaller scales. This then leads to a higher abundance of more
massive clusters. Figure~\ref{fig:7} illustrates these effects.

The peak in the mass distribution shifts to lower masses as the baryon
fraction or the Hubble constant increase. While the amplitude of the
exponential cut-off is unchanged while $h$ is kept fixed, it increases
when $h$ decreases, as expected. The most prominent effect is the
change in the peak height with baryon fraction.

\subsection{Mass determination}

The simple relation between $M_\mathrm{ap}$ and the cluster mass $M$
derived earlier and shown in Fig.~\ref{fig:3b} can now be used to
derive cluster masses. Equations~(\ref{eq:3.7}) and (\ref{eq:3.10})
imply
\begin{equation}
  M_\mathrm{ap}=\left(\frac{3}{800\pi}\right)^{1/3}\,
  \frac{\delta_\mathrm{c}}{c}\,
  \frac{\rho_\mathrm{crit}^{2/3}}{\Sigma_\mathrm{cr}}\,
  M^{1/3}\equiv C_M\,M_{15}^{1/3}\;,
\label{eq:4.1}
\end{equation}
where $r_\mathrm{s}=r_{200}/c$ and the definition (\ref{eq:0.0}) of
$r_{200}$ have been inserted. Because of the fairly narrow redshift
range occupied by the cluster sample, and the low redshift where they
are typically located, $\Sigma_\mathrm{cr}$ can be considered constant
in redshift. What is more, $C_M$ in (\ref{eq:4.1}) is almost
independent of cosmological parameters. Changing $\Omega_0$ from $0.3$
to $1$ changes $C_M$ by less than 5 per cent. For an aperture radius
of $1'$, we find
\begin{equation}
  M_\mathrm{ap}\approx(0.18\pm0.01)\,M_{15}^{1/3}\;,
\label{eq:4.2}
\end{equation}
and this relation continues to hold even for moderatly high cluster
redshifts. It is therefore straightforward to determine the absolute
masses of the clusters in the combined gravitationally lensing
Sunyaev-Zel'dovich cluster sample.

\begin{figure}[ht]
  \centerline{\includegraphics[width=\fsize]{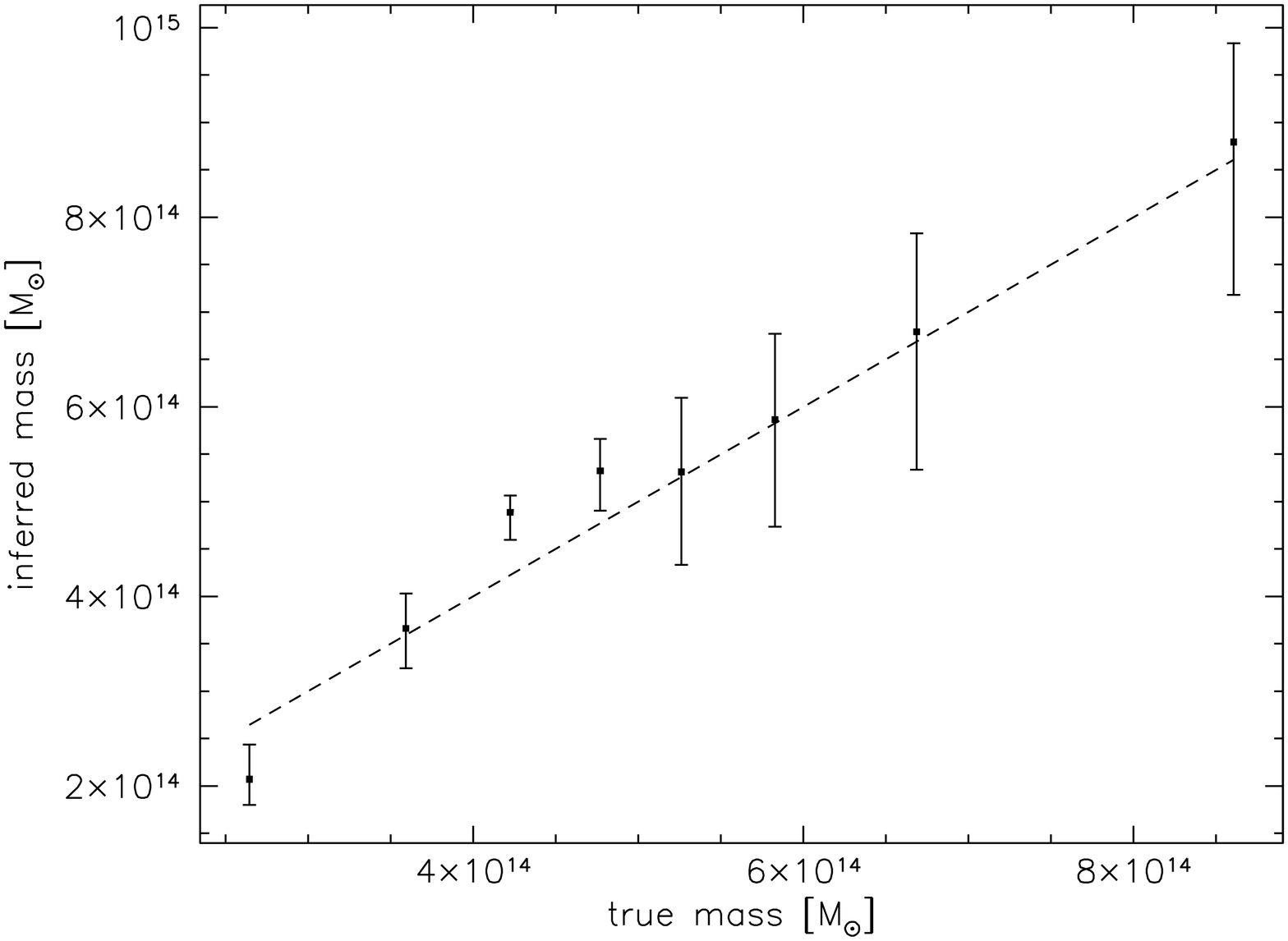}}
\caption{Comparison of inferred and true masses of clusters drawn
randomly from their expected mass-redshift distribution in a
low-density, flat, cluster-normalised CDM model universe. The aperture
mass $M_\mathrm{ap}(1')$ on an angular scale of $1'$ was calculated
and converted to mass via eq.~(\ref{eq:4.2}). Each data point shows
the median mass derived from an equal number of clusters. Error bars
show the $\pm34\%$ deviation from the mean, i.e.~they would correspond
to 1-$\sigma$ error bars if the distribution was Gaussian.}
\label{fig:8}
\end{figure}

In order to test the accuracy of this mass determination, we have
randomly drawn a set of clusters from the number-density distribution
shown in Fig.~\ref{fig:4}, with the total number computed for a flat,
low-density, cluster normalised CDM universe. For each of these
clusters, we computed the aperture mass $M_\mathrm{ap}(1')$ for an
aperture radius of $1'$, and then used Eq.~(\ref{eq:4.2}) to infer its
virial mass. Figure~\ref{fig:8} compares the inferred and true masses,
binned in such a way that each bin contains the same number of
clusters.

\begin{figure}[ht]
  \centerline{\includegraphics[width=\fsize]{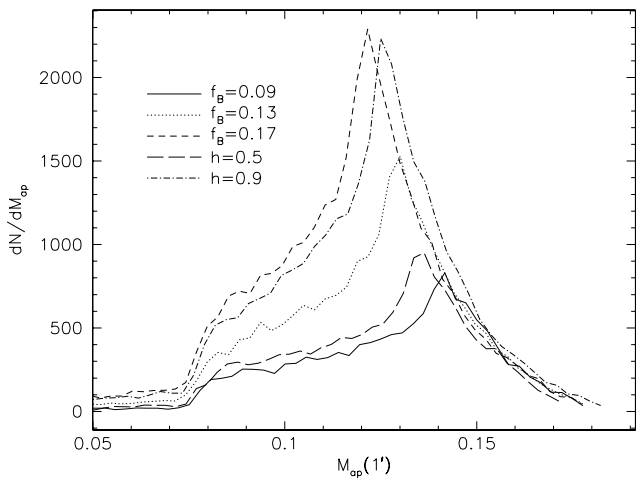}}
\caption{Similar to Fig.~\ref{fig:7}, this figure shows the
distribution of clusters as a function of $M_\mathrm{ap}$ rather than
mass. Five curves are shown to illustrate the effects of changing
baryon fraction and Hubble constant, as indicated.}
\label{fig:9}
\end{figure}

Data points show the median of the inferred masses in each bin, and
error bars enclose the $\pm34\%$ deviation from the median, i.e.~they
loosely correspond to 1-$\sigma$ error bars. As Fig.~\ref{fig:8}
shows, the accuracy of the inferred cluster masses is of order
$20\%$. The distribution of clusters as a function of $M_\mathrm{ap}$
prominently reflects the peak in the mass distribution, as
Fig.~\ref{fig:9} shows.

Using (\ref{eq:4.2}), the true mass corresponding to the peak location
in the mass function can accurately be determined from the peak in the
$M_\mathrm{ap}$ distribution. The $M_\mathrm{ap}$ distribution shows
the same behaviour in reaction to changes in baryon fraction and
Hubble constant as the mass distribution shown in Fig.~\ref{fig:7}. In
particular, the peak height changes substantially with the baryon
fraction, and much less so with the Hubble constant.

Relation (\ref{eq:4.2}) holds for clusters with an NFW density
profile. It is valid because the aperture-mass profile is essentially
flat for small aperture sizes, so that a weak-lensing estimate inside
an area with fixed angular size can be used to extract a global halo
property. Generally, the cluster mass profile needs to be known in
order to relate aperture masses to virial masses. Surface-density
profiles of clusters can be constrained with the same weak-lensing
data used to determine the aperture masses, so that masses can be
determined from weak lensing even if the true profile deviates from
the NFW form.

However, the conversion from aperture mass to virial mass is not
necessary for the extraction of cosmological information from the
combined gravitationally lensing Sunyaev-Zel'dovich cluster
sample. Rather, one can use weak-lensing quantities to classify
clusters, and then use direct weak-lensing simulations to compare
models to data.

\section{Summary\label{sec:5}}

Full-sky microwave surveys like the upcoming {\em Planck\/} mission
will detect of order $10^4$ galaxy clusters through their thermal
Sunyaev-Zel'dovich effect. Using modified Press-Schechter theory in
CDM model universes normalised to the local abundance of rich
clusters, and scaling relations derived from the spherical collapse
model, we have investigated the physical properties which that cluster
sample is expected to have. We have then addressed the question what
additional information can be gained by weak-lensing follow-up
observations of the clusters in the Sunyaev-Zel'dovich sample. For
doing so, we described individual clusters with the density profile of
Navarro et al.~(1997) and applied the aperture mass as a measure for
their weak-lensing effects. Our main results can be summarised as
follows:

\begin{itemize}

\item Taking beam convolution and background fluctuations into
account, the number of Sunyaev-Zel'dovich clusters detectable for {\em
Planck\/} is a few times $10^4$ if a baryon fraction of $13\%$ is kept
fixed, quite independent of the cosmological parameters. If the baryon
fraction is adapted to the primordial nucleosynthesis constraint, it
drops with increasing $\Omega_0$, and the number of detectable
Sunyaev-Zel'dovich clusters falls by a factor of $\sim3$ as $\Omega_0$
increases from $0.3$ to $1$.

\item One particular measure of weak gravitational lensing, the
aperture mass, effectively determines the convergence scale
$\kappa_\mathrm{s}$ for a cluster described by an NFW density profile,
and the Einstein radius for a singular isothermal sphere. The
aperture-mass profile of clusters depends sensitively on the density
profile. While it falls with aperture size $\theta$ as $\theta^{-1}$
for singular isothermal spheres, it is much flatter for a NFW haloes.

\item The significantly lensing sub-sample comprises more than $70\%$
of the original Sunyaev-Zel'dovich cluster sample. The combined sample
is narrowly confined in redshift and mass. The majority of redshifts
fall within $0.2\pm0.1$, and the masses within
$(5\pm3)\times10^{14}\,M_\odot$. Approximate analytic descriptions of
the sample's number density in the $M$-$z$ plane are given.

\item Because of the narrow redshift range and the generally low
cluster redshifts, the weak-lensing effects are virtually independent
of the source redshift distribution. Under these circumstances, the
aperture mass essentially measures the radial scale of the NFW density
profile, which can easily be converted to virial mass. The relation
$M_\mathrm{ap}\propto M^{1/3}$ holds to very good
approximation. Cluster masses can be determined with that relation
with an accuracy of $\sim20\%$.

\item The mass distribution of the combined cluster sample exhibits a
sharp low-mass cutoff and an exponential decrease at high masses. For
low-density universes, the mass distribution has a sharp peak near
$5\times10^{14}\,M_\odot$. The low-mass cutoff is due to the lack of
significant lensing effects in low-mass Sunyaev-Zel'dovich clusters,
since they have low redshifts. The high-mass fall-off reflects the
exponential decline in the cluster mass function. The peak occurs at
the minimum mass required for a cluster to be a detectable
Sunyaev-Zel'dovich source at all redshifts. This is possible because
the squared angular diameter distance, divided by $(1+z)$, has a
maximum at a finite redshift $z_1\la1$. The peak occurs if clusters
exist at such redshifts, i.e.~in low-density universes.

\item The cluster mass at the peak location can accurately be measured
if the cluster density profile is known. In combination with the
exponential high-mass cut-off of the mass distribution, the peak
location allows the determination of the amplitude of the dark-matter
power spectrum and the baryon fraction in the intracluster gas.

\end{itemize}

Of course, we rely on the validity of Press-Schechter theory, as
modified by Sheth \& Tormen (1999) and Jenkins et al.~(2000). The mass
function derived by Jenkins et al.~from numerical simulations agrees
very well with the analytical expression by Sheth \& Tormen, showing
that there are good reasons to believe that the mass function used
here approximates reality sufficiently closely.

Although all our results were derived under the assumption that
clusters have the density profile found by Navarro et al.~(1997), this
assumption is not critical for the results. The property shown, that
the aperture mass becomes independent of aperture size for small
apertures, is convenient for our purpose in that it facilitates the
conversion of aperture masses to virial masses, but by no means
necessary. The NFW density profile has recently been disputed by
studies finding a steeper central density cusp (e.g.~Moore et
al. 1999), albeit on radial scales much smaller than probed by
weak-lensing techniques. Weak lensing observations will allow the
determination of the mass profile. The fact that the combined
gravitationally lensing Sunyaev-Zel'dovich cluster sample is confined
to a fairly narrow, low-redshift range implies that weak-lensing
results become essentially independent of the source-redshift
distribution.

It therefore appears safe to conclude that cluster samples selected by
their thermal Sunyaev-Zel'dovich effect and combined with weak-lensing
follow-up observations, are well confined in redshift and mass, and
provide a unique opportunity to accurately measure the cluster mass
function at redshifts around $0.2$, thus the amplitude of the
dark-matter power spectrum, and the baryon fraction in the
intracluster gas.

The most prominent feature in the mass function of the combined
gravitationally lensing, Sunyaev-Zel'dovich cluster sample is the
pronounced peak predicted to occur for low-density universes. Its
physical origin is the flattening of the relation between
angular-diameter distance and redshift away from Hubble's law due to
space-time curvature, and the thermal history of the cluster
population, for which the spherical collapse model predicts that
temperature rises as $(1+z)$ for fixed cluster mass. Adopting this
model, we saw that location and height of the peak provide additional
information on the intracluster baryon fraction. It is highly likely
that the microwave background observations leading to the compilation
of the Sunyaev-Zel'dovich cluster sample will allow accurate
determinations of the cosmological parameters including the baryon
fraction (e.g.~Efstathiou \& Bond 1999). The gravitational-lensing
follow-up observations will then supply valuable tests for the
consistency between the detected and expected cluster populations, and
additional information on the thermal history of the cluster
population.

Finally, it appears recommendable to avoid the conversion from
weak-lensing measures like the aperture mass to virial mass. It is
equally possible to define a cluster sample not by its mass, but
operationally by the integrated Sunyaev-Zel'dovich decrement or
increment and weak-lensing measures like the aperture mass. Comparing
Figs.~\ref{fig:7} and \ref{fig:9} shows that the aperture-mass
distribution exhibits the same features as the mass distribution
itself. Uncertainties in the conversion between lensing effects and
masses could be avoided if measured aperture-mass distributions would
directly be compared to simulations such as those leading to the
results shown in Fig.~\ref{fig:9}.

\section*{Acknowledgements}

I wish to thank Adi Nusser, Peter Schneider and Saleem Zaroubi for
enlightening discussions, and Simon White for detailed comments on the
manuscript.

\end{document}